\documentclass[aip,reprint]{revtex4-2}
\usepackage[T1]{fontenc}
\usepackage[utf8]{inputenc}
\usepackage[russian,english]{babel}
\usepackage{graphicx}
\usepackage{amsmath}
\usepackage{amssymb}
\usepackage{physics}
\usepackage{color,soul}
\usepackage{todonotes}
\usepackage{hyperref}
\usepackage{cancel}

\DeclareMathAlphabet\mathbfcal{OMS}{cmsy}{b}{n}

\graphicspath{{./Figs/}}

\begin{document}

\title{Hydrodynamical model of QED cascade expansion in an extremely strong laser pulse}

\date{\today}
\author{A.~S.~Samsonov}
\email{asams@ipfran.ru}
\author{I.~Yu.~Kostyukov}
\author{E.~N.~Nerush}
\affiliation{Institute of Applied Physics of the Russian Academy of
Sciences, 46 Ulyanov St., Nizhny Novgorod 603950, Russia}

\begin{abstract}
    Development of the self-sustained quantum-electrodynamical (QED) cascade in a single strong laser pulse is studied analytically and numerically. The hydrodynamical approach is used to construct the analytical model of the cascade evolution, which includes the key features of the cascade observed in 3D QED particle-in-cell (QED-PIC) simulations such as the magnetic field predominance in the cascade plasma and laser energy absorption. The equations of the model are derived in the closed form and are solved numerically. Direct comparison between the solutions of the model equations and 3D QED-PIC simulations shows that our model is able to describe the complex nonlinear process of the cascade development qualitatively well. The various regimes of the interaction based on the intensity of the laser pulse are revealed in both the solutions of the model equations and the results of the QED-PIC simulations.
\end{abstract}

\maketitle

\section{Introduction}
\label{sec.Introduction}
There is strong evidence that the development of QED cascades is an immanent feature of interaction of an extremely strong electromagnetic fields with matter in a majority of configurations~\cite{sturrock1971model,daugherty1982electromagnetic,nerush2007radiation,Bell2008,Nerush2011,Ridgers2012,narozhny2015quantum,Kostyukov2016,nerush2017weibel,efimenko2019laser,yakimenko2019prospect}. The essence of the cascade is emission of the high-energy photons by the ultrarelativistic particles (the nonlinear Compton scattering) and the subsequent decay of these photons into the electron-positron pairs (the Breit-Wheeler process) which leads to the multiplication of particles. These processes are believed to play an important role in many astrophysical phenomena like cosmic ray showers~\cite{bhabha1937passage}, processes in pulser magnetosphere~\cite{sturrock1971model,ruderman1975theory,daugherty1982electromagnetic},  $\gamma$-ray bursts~\cite{meszaros2006gamma} etc. 
The variety and complexity of the $e^+e^-$ plasma structures produced in QED cascading makes it clear that there is no simple way to tackle the problem. One of the reasons behind it is that the emission of the photons by the electrons and the positrons greatly alters the dynamics of the latter~---~the effect known as the radiation reaction. Accurate description of the radiation reaction is a long standing problem of both classical and quantum electrodynamics~\cite{Zeldovich,rohrlich2007classical,ritus1985quantum,BaierKatkov}.

With the upcoming multi-PW level laser facilities such as ELI~\cite{ELI} and Apollon~\cite{Apollon} these processes are expected to be observed in the laboratory light-matter interaction experiments. An extensive search for the optimal configuration of such experiments is being conducted nowadays~\cite{gonoskov2013probing,gelfer2015optimized,grismayer2016laser,grismayer2017seeded,jirka2017qed,luo2018qed,Zhang2020}. 
As already mentioned the QED cascade is a significantly nonlinear phenomenon that is why its analytical study is complicated. And while PIC simulations serve as a starting point for most of the theoretical research and can give a valuable insight into the nature of the discussed processes, even deriving phenomenological laws or scaling can be extremely time consuming as it usually requires scanning over the multi-dimensional map of the parameters. These dependencies though are crucial to design the experiments with the next generation of laser facilities. 
Various schemes are proposed in order to lower the threshold of QED cascading: multiple laser pulses with small number of seed particles, laser-beam interaction, etc.~\cite{nerush2011analytical,Kostyukov2016,grismayer2017seeded,jirka2017qed,del2017ion,kostyukov2018growth,Yuan2018,Luo2018,Lu2018,Martinez2019}
The key reason for such configurations to be optimal is that the governing parameter of the QED processes $\chi$ which is the ratio of the transverse component of the effective electric field experienced by the relativistic particle in the rest frame to the critical Sauter-Schwinger field~\cite{berestetskii1982quantum} is maximaized in these scenarios. There is a crucial difference though between these two approaches. In the first case the cascade gains its energy from the electromangetic field and this is so-called A-type cascade, while in the second one the cascade energy is mostly limited by the initial energy of the seed and this is called the S-type cascade~\cite{mironov2014collapse}.

\begin{figure}
    \includegraphics[width=85mm]{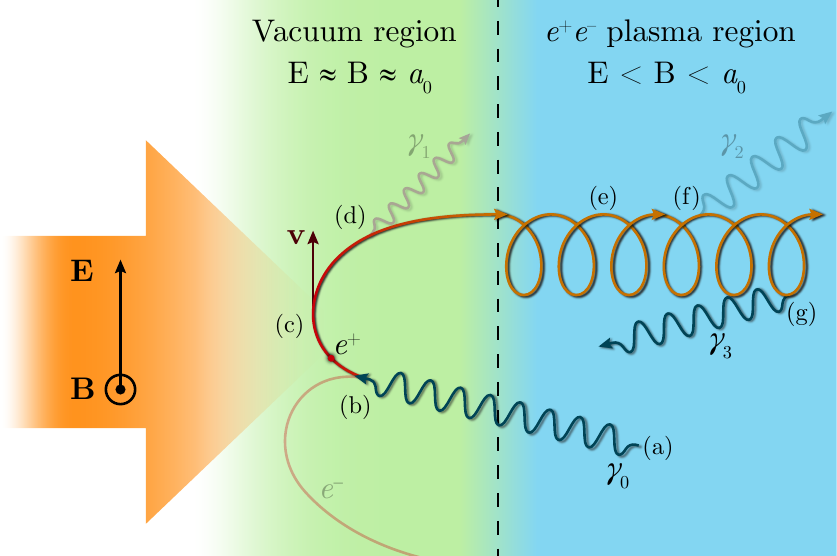}
    \caption{\label{fig.scheme} Core mechanism of the cascade self-sustenance. (a), (f), (g) Emission of the gamma-quant in the plasma region or the \textit{involved} gamma-quant, (b) the \textit{involved} gamma-quant decay in the vacuum region, (c) positron and electron acceleration in the plane wave, (d) emission of the gamma-quant in the vacuum region or the \textit{decoupled} gamma-quant, (e) helical motion of the positron in the plasma region.}
\end{figure}

A specific configuration of QED cascade development which occurs in the interaction of the extremely intense laser pulse with the solid target is recently explored by 3D QED-PIC simulations~\cite{Samsonov2019}. The peculiar mechanism of that cascade development makes it difficult to attribute it to either of A-type or S-type cascade. 
For the sake of convenience we will briefly describe the core mechanism of the discussed cascade (see Fig.~\ref{fig.scheme} for the visual schematics of the process). The main feature that allows the cascade to sustain itself is the fact that the collective motion of the electrons and the positrons alters the laser field so that its magnetic component becomes larger than the electric one while staying mutually nearly perpendicular. In such fields the particles drift along the direction of the laser propagation with the drift velocity and rotate in the plane perpendicular to the direction of the magnetic field, so their trajectories are helical (see Supplemental material~\cite{Supplemental}). The particles radiate gamma-quanta along the direction of the instant tangent to the trajectory. Occasionally that direction may be opposite to the direction of the laser propagation. Such gamma-quanta eventually leave the electron-positron plasma and get to the vacuum where strong laser field is present and where they are highly probable to photoproduce new electron-positron pairs. The newly created pair is then accelerated by the laser pulse and pushed towards the plasma region and the process repeats. 
As a results the QED cascade continuously expands towards the laser pulse building a pairs plasma `cushion'. This proccess is similar to the propagation of the ionization front in the microwave gas discharge~\cite{semenov1982breakdown}

The models describing electrodynamics of the cushion plasma~\cite{kirk2013pair,Samsonov2019} and QED cascade evolution~\cite{Samsonov2019} are proposed. However these models are not self-consistent. The model proposed in Ref.~\cite{kirk2013pair} does not take into account particle multiplication because of cascading while the model in Ref.~\cite{Samsonov2019} neglects the laser field depletion because of absorption in the cascade plasma. In this paper we construct the self-consistent model which describes spatio-temporal evolution of both the laser field and the cascade plasma. The goal of this paper is twofold. First, we develop a reduced model which requires much less computational resources than that for 3D QED-PIC simulations. Second the model development and its verification allows us to understand and evaluate the role of different physical processes behind QED cascading in a single laser pulse.

The paper is organized as follows. In Sec.~\ref{sec.Main} we formulate the problem and start from the kinetic equations including the QED processes. Next we propose a set of general simplifications which allow to greatly reduce the complexity of the equations. According to these simplifications self-consistent hydrodynamics equations are derived. In Sec.~\ref{sec.Discussion} the method of the numeric solution of the derived equations is discussed and the results of the solution are compared to the results of the QED-PIC simulations. The obtained results are summarized and discussed in Sec.~\ref{sec.Conclusion}. In Appendix~\ref{app.Acceleration} we examine the problem of the electron acceleration in the plane wave and in Appendix~\ref{app.Electrodynamics} we derive the electrodynamical properties of the dense electron-positron plasma. Results of both these problems are used in the model equations derivation.

\section{QED cascade development model derivation}
\label{sec.Main}
Similar to Refs.~\cite{nerush2011analytical, elkina2011qed} we start our analysis from the kinetic equations for electrons, positrons and gamma-quanta, assuming that the QED cascade is in the self-sustaining stage so the seed particles (e.g. the electrons and the ions of the target) do not contribute to it. The kinetic equations and the Maxwell's equations can be written as follows
\begin{align}
    \label{eq.Boltzman1}
    \frac{\partial f_{e^\pm}}{\partial t} +  \mathbf{v}_{e^\pm} \nabla f_{e^\pm} \pm \left( \mathbf{E} + [\mathbf{v}_{e^\pm} \times \mathbf{B}] \right) \frac{\partial f_{e^\pm}}{\partial \mathbf{p}} = &  \\ \nonumber
    \begin{aligned}
        = &\int f_\gamma(\mathbf{p'}) w_{pair}(\mathbf{p', p}) d\mathbf{p'} + \\ \nonumber
        + &\int f_{e^\pm}(\mathbf{p'}) w_{rad}(\mathbf{p', p}) d\mathbf{p'} - \\ \nonumber 
        - &\int f_{e^\pm}(\mathbf{p}) w_{rad}(\mathbf{p, p'}) d\mathbf{p'}  ,
    \end{aligned} & 
\end{align}
\begin{align}
    \label{eq.Boltzman2}
    \frac{\partial f_\gamma}{\partial t} + \mathbf{v}_\gamma\nabla f_\gamma = &\int  f_{e^\pm}(\mathbf{p'}) w_{rad}(\mathbf{p', p'-p}) d\mathbf{p'} - \\ \nonumber
    + &\int f_\gamma(\mathbf{p}) w_{pair}(\mathbf{p, p'}) d\mathbf{p'} ,
\end{align}
\begin{align}
    \label{eq.Max1}
    \nabla\times \mathbf{E} &= -\frac{\partial \mathbf{B}}{\partial t}, \\
    \label{eq.Max2}
    \nabla\times \mathbf{B} &= \frac{\partial \mathbf{E}}{\partial t} + \int f_{e^+} \mathbf{v}_{e^+} d\mathbf{p} - \int f_{e^-} \mathbf{v}_{e^-} d\mathbf{p} ,
    \end{align}
where $f_{e^\pm,\gamma}(t,\mathbf{r,p})$ are the distribution functions of the electrons, the positrons and the gamma-quanta respectively, $\mathbf v$ is the particle velocity which is equal to $\mathbf{p}/\sqrt{1+p^2}$ for the electrons and the positrons and to $\mathbf{p}/p$ for the gamma-quanta, $w_{rad}(\mathbf{p', p})d\mathbf{p'}$ is the probability in the time unit for the electron or the positron with the momentum $\mathbf{p'}$ to emit the gamma-quant with the momentum $\mathbf{p'-p}$, $w_{pair}(\mathbf{p', p})d\mathbf{p'}$ is the probability in the time unit for a gamma-quant with the momentum $\mathbf{p'}$ to photoproduce the electron with the momentum $\mathbf{p}$ and the positron with the momentum $\mathbf{p'-p}$. 
We use a common relativistic normalization where the electric and the magnetic fields are normalized to the value of $m_ec\omega_L/e$, where $m_e$ and $e>0$ are the mass and the charge of the electron, $c$ is the speed of light and $\omega_L$ is the characteristic frequency of the external field, the particle number densities are normalized to the critical density $n_c=m_e\omega_L^2/4\pi e^2$, the energies and the momenta are normalized to $m_e c^2$ and $m_e c$ respectively, the coordinates and the time are normalized to $c/\omega_L$ and $1/\omega_L$ respectively, thus the velocities are normalized to the $c$.

\subsection{Model assumptions}
Let us apply several assumptions to simplify the model. First of all, as we investigate interaction with the plane EM wave, the problem can be considered spatially one-dimensional. If we also restrict ourselves to the circularly-polarized laser pulses then the symmetry relative to the rotation along the axis of the pulse propagation can be also utilized. This simplifications lead to the distribution functions becoming dependant on three variables (excluding time) rather than six: $f(t;\mathbf{r},\mathbf{p})=f(t;x,p,\theta)/2\pi$, where $p$ is the momentum of the particle and $\theta$ is the angle between the momentum and the $x$-axis.

\begin{figure}
    \includegraphics[width=80mm]{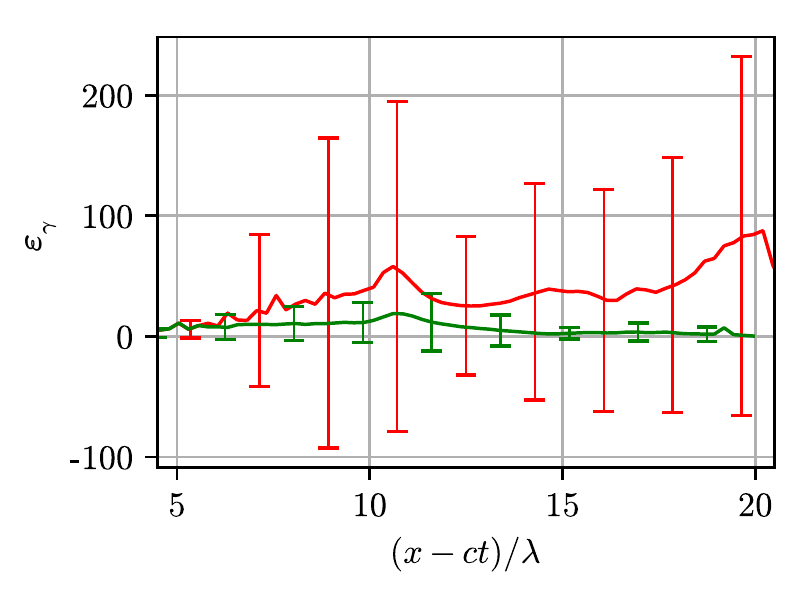}
    \caption{\label{fig.gamma_energy} The mean energy $\varepsilon_\gamma$ of the gamma-quanta located in the vicinity of the $x$-coordinate calculated from all the particles (red line) and from the particles with the velocity along the $x$-axis less than $0.5$ which supposedly include only the \textit{involved} gamma-quanta (green line). The error bars depict the standard deviation. The data is taken from the results of the PIC simulation for the time instance $ct/\lambda=18$. The simulation parameters are discussed in Sec~\ref{sec.Discussion}. The initial conditions are the same as in Fig~\ref{fig.sol2500}.}
\end{figure}

Secondly, we will assume all distribution functions to be locally monoenergetic, i.e. $f\propto\delta(p-\overline{p}(x))/p^2$, where $\overline{p}(x)$ is the mean value of the momentum of the particles located in the small vicinity of $x$. We denote the mean energy of gamma-quanta as $\varepsilon_\gamma$ and the mean energy of pairs as $\varepsilon_p$, assuming that particles are ultrarelativistic and thus $\varepsilon_p^2 = 1+p_p^2\approx p_p^2$. While the monoenergetic approximation is quite strong, we suppose that the mechanism of the cascade development explained in Sec.~\ref{sec.Introduction} does not rely on any particular feature of the particles energetic spectra. So we argue that accounting energetic spectra evolution in our model will cause only quantitative changes rather than qualitative ones while greatly complicating the equations. It will be discussed later that this assumption is valid for the pairs which enter the plasma region with approximately equal energies. 
As for gamma-quanta we actually use a two-stream approximation, i.e. we separate gamma-quanta into ones that are emitted in the vacuum region and propagate mostly along the direction of the laser pulse propagation ($x$-axis) and thus do not contribute to the cascade (we designate them as \textit{decoupled} gamma-quanta) and ones that are emitted in the plasma region in many different directions and provide a positive feedback needed for the cascade development (we designate them as either \textit{involved} gamma-quanta or simply gamma-quanta). As Fig.~\ref{fig.gamma_energy} shows the energy spectrum of the gamma-quanta is broad, while if we exclude the \textit{decoupled} gamma-quanta, then the width of the spectrum diminishes significantly which justifies our assumption. As the \textit{decoupled} gamma-quanta affect the cascade development only by taking away some portion of the total energy, their spatial distribution is irrelevant for the cascade but it will be calculated for more explicit comparison with the results of the QED-PIC simulations.

In order to omit integration over energies and azimuth angle $\varphi$ we redefine $f$ as follows
\begin{equation}
    \begin{split}
        f(x, \varepsilon, \theta, \varphi) \rightarrow \int\limits_0^\infty\int\limits_0^{2\pi}f(x,\varepsilon,\theta,\varphi)2\pi\varepsilon^2d\varphi d\varepsilon =\\
        = n(x)\Phi(\theta).
    \end{split}
\end{equation}
where $n(x)$ is the particles density distribution, and $\Phi(\theta)$ is the particles momenta angular distribution such that
\begin{align}
    \int\limits_{-\infty}^{+\infty} n(x)dx = N, \\
    \int\limits_0^\pi \Phi(\theta)\sin\theta d\theta = 1.
\end{align}
where $N$ is the total number of particles.
Assuming monoenergetic distribution hints that we can use a hydrodynamical approach by calculating moments of the distribution functions from Eqs.\eqref{eq.Boltzman1}--\eqref{eq.Boltzman2}. The following quantities can be introduced
\begin{align}
    \label{eq.Wpair0}
    W_{pair}(\chi_\gamma, \varepsilon_\gamma) = \int w_{pair}(\mathbf{p},\mathbf{p'})d\mathbf{p'} ,\\
    \label{eq.Wrad0}
    W_{rad}(\chi_p, \varepsilon_p) = \int w_{rad}(\mathbf{p},\mathbf{p'})d\mathbf{p'} ,\\
    \label{eq.Irad0}
    I_{rad}(\chi_p) = \int w_{rad}(\mathbf{p},\mathbf{p'})(\varepsilon_p-\varepsilon_p')d\mathbf{p'},
\end{align}
where $W_{pair}, W_{rad}, I_{rad}$ are the total probabilities of the pair photoproduction (or gamma-quant decay), gamma-quanta emission and the intensity of the gamma radiation, respectively~\cite{BaierKatkov}. Note that these quantities depend on the Lorentz-invariant QED parameter $\chi$
\begin{equation}
    \chi = \frac{\varepsilon}{\mathcal{E}_S} \sqrt{ \left(\mathbf{E}+\mathbf{v}\times\mathbf{B}\right)^{2}-\left(\mathbf{v}\cdot\mathbf{E}\right)^{2} } ,
\end{equation}
where $\varepsilon$ is the particle energy, $\mathcal{E}_S = eE_S/m_ec\omega_L = m_ec^2/\hbar\omega_L$ and  $E_S={m_e}^2c^3/\hbar e$ is the critical Sauter-Schwinger field~\cite{berestetskii1982quantum}.

Generally the hydrodynamics equations have form of a continuity equation, i.e.
\begin{equation}
    \frac{\partial D_\alpha}{\partial t} + \frac{\partial F_\alpha}{\partial x} = \sum_{\beta} S[\alpha, \beta] ,
\end{equation}
where $D_\alpha$ and $F_\alpha$ are the density and the flux of some value $\alpha$ and $S[\alpha,\beta]$ is the source responsible for change of the value $\alpha$ in the process $\beta$. 
Note that despite the fact that we specified the energy distribution of the particles, to calculate the sources $S[\alpha,\beta]$ one also needs to know the angular distribution of the particle which will be discussed below.
We suppose that the following set of the equations can quantitatively describe the cascade development
\begin{widetext}
\begin{align}
    \label{eq.hd_p1}
    \frac{\partial }{\partial t}n_p  +\frac{\partial}{\partial x}\left( v_x n_p \right) &= S[n, pp] ,\\
    \label{eq.hd_p2}
    \frac{\partial}{\partial t}\left( \varepsilon_p n_p \right)  + \frac{\partial}{\partial x}\left( v_x \varepsilon_p n_p \right)  &=  S[\varepsilon, pp] + S[\varepsilon, acc] \psi_{vac} - S[\varepsilon, rad_c] \psi_{pl} - S[\varepsilon, rad_d] \psi_{vac} ,\\
    \label{eq.hd_g1}
    \frac{\partial }{\partial t}n_\gamma  +\frac{\partial}{\partial x}\left( \overline{v_\gamma} n_\gamma \right)&=-S[n, pp] + 2 S[n, rad_c] \psi_{pl} ,\\
    \label{eq.hd_g2}
    \frac{\partial}{\partial t}\left( \overline{v_\gamma} n_\gamma \right) + \frac{\partial}{\partial x}\left( \overline{v_\gamma^2} n_\gamma \right) &= -S[v, pp] + 2 S[v, rad_c] \psi_{pl} ,\\
    \label{eq.hd_g3}
    \frac{\partial}{\partial t}\left( \varepsilon_\gamma n_\gamma \right) +\frac{\partial}{\partial x}\left( \overline{v_\gamma} \varepsilon_\gamma n_\gamma \right) &= -S[\varepsilon, pp] + 2 S[\varepsilon, rad_c] \psi_{pl} ,\\
    \label{eq.hd_e1}
    \frac{\partial }{\partial t}\left( \frac{E^2 + B^2}{2} \right) +\frac{\partial}{\partial x} {\left[ \mathbf{E\times B} \right]}_x &= - 2 S[\varepsilon, acc]\psi_{vac} ,\\
    \frac{\partial \Sigma_\gamma}{\partial t} &= 2 \int\limits_0^\infty S[\varepsilon, rad_d]\psi_{vac} dx ,
\end{align}
\end{widetext}
where $n_p = n_{e^+} = n_{e^-}$ is half the density of the electron-positron plasma assuming that cascade plasma is quasineutral, $v_x$ is the mean velocity of the pairs which is calculated below and $\overline{v_\gamma}$ and $\overline{v_\gamma^2}$ are the mean longitudinal velocity and mean square longitudinal velocity of the gamma-quanta which are calculated from their angular distribution as follows
\begin{align}
    \overline{v_\gamma} = \int_0^\pi \Phi(\theta)\cos\theta\sin\theta d\theta, \\
    \overline{v_\gamma^2} = \int_0^\pi \Phi(\theta)\cos^2\theta\sin\theta d\theta.
\end{align}
Eq.~\eqref{eq.hd_e1} is the Poynting's theorem which is essentially the continuity equation for the electromagnetic energy. The sources $S[n,\beta]$, $S[v,\beta]$ and $S[\varepsilon,\beta]$ correspond to the change in the particle densities, the longitudinal velocity and the energy respectively, the sources $S[\alpha, pp]$, $S[\alpha, acc]$, $S[\alpha, rad_c]$ and $S[\alpha, rad_d]$ correspond to the processes of the pairs photoproduction, pairs acceleration in the electromangetic field, \textit{involved} gamma-quanta emission by the pairs in the plasma region and \textit{decoupled} gamma-quanta emission by the pairs in the vacuum region respectively [labeled (b), (c), (d) and (f) respectively in Fig.~\ref{fig.scheme}]. $\Sigma_\gamma$ is the total energy of the \textit{decoupled} gamma-quanta. The factor $\psi_{vac}(\psi_{pl})$ is equal to 1 in the vacuum (plasma) region and is equal to 0 in the plasma (vacuum) region. These factors will be specified below. Note that $\psi_{vac}+\psi_{pl}=1$. For the sake of convenience these factors will be omitted where it is clear which region is being discussed.

\subsection{Electromagnetic field configuration}
\label{sub.fields}
In the vacuum region the fields are close to the plane wave, i.e. the magnitudes of the electric and the magnetic fields are equal $E\approx B$ and they are mutually perpendicular $\mathbf{E\cdot B}\approx 0$. According to the results of the 3D QED-PIC simulations the electric and the magnetic fields inside the plasma region stay almost mutually perpendicular and magnetic field magnitude is everywhere larger than the electric one $B > E$. The spatial distribution of the electromagnetic field has a characteristic scale of $\lambda$ in both regions. 
In such field configuration the charged particle drifts perpendicular to the both the electric and the magnetic field with the velocity $\mathbf{v_d}=[\mathbf{E\times B}]/B^2$. Assuming that the laser pulse propagates along the $x$-axis and thus the fields lay in the $yz$-plane this velocity is directed along the $x$-axis. Assuming that the fields are mutually perpendicular we get
\begin{equation}
    \label{eq.vdrift}
    v_x\approx E/B.
\end{equation}
In the vacuum region the particles move in the intense EM plane wave with equal electric and magnetic fields $E=B$. If the initial energy of the particle $\varepsilon$ is smaller than the field amplitude $E$ then on the timescale much smaller than the laser period the particle longitudinal velocity tends to the speed of light, i.e. $v_x\approx 1=E/B$ (see Appendix~\ref{app.Acceleration}). So we assume that Eq.~\eqref{eq.vdrift} is valid in both the vacuum and the plasma region.

\subsection{Coupled gamma-quanta distribution function}
\label{sub.gammas}
\begin{figure*}
    \includegraphics[width=175mm]{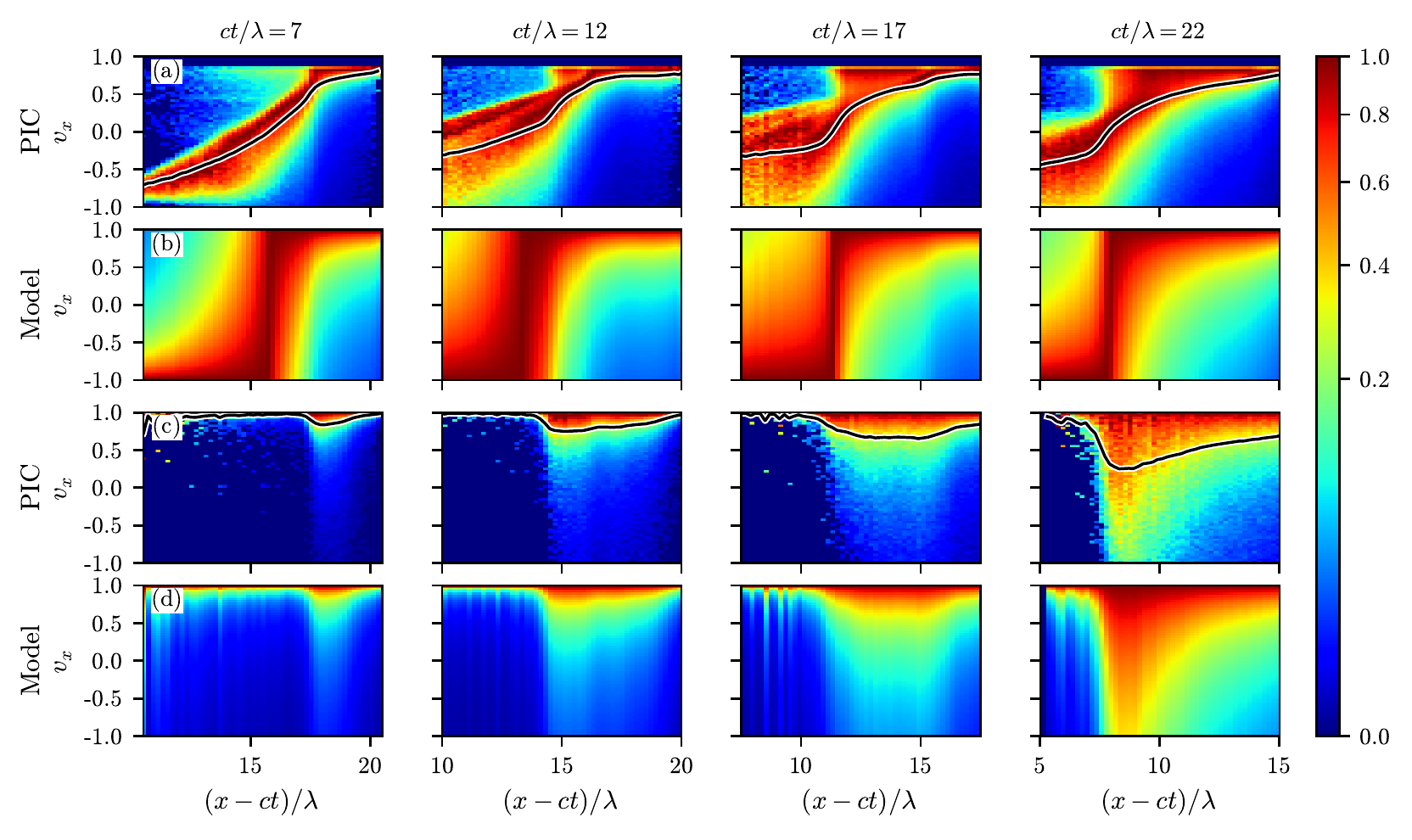}
    \caption{\label{fig.ang} Validation of the approximation used for describing angular distribution of the particles. The angular distribution of the particles [(a)~---~gamma-quanta, (c)~---~pairs] located in the vicinity of the coordinate $x$ (color map) and the mean longitudinal velocity calculated from the distribution (black line) computed from the results of the QED-PIC simulation. The angular distribution calculated from the mean velocity using the expression~\eqref{eq.Phi} for gamma-quanta (b) and pairs (d). }
\end{figure*}

As discussed in Sec.~\ref{sec.Introduction} the \textit{involved} gamma-quanta are emitted by the pairs during their motion along the helical trajectories in the plasma region (see also Supplemental material~\cite{Supplemental}). Because of that the angular distribution of the gamma-quanta is quite wide. We will assume that it is smooth and can be described by the single parameter. This parameter is the velocity of the instantaneous reference frame in which the angular distribution of the gamma-quanta located in the vicinity of the $x$-coordinate is uniform. Thus in the laboratory reference frame the total distribution function of the \textit{involved} gamma-quanta has the following form
\begin{align}
    f_\gamma(t; x,\theta)=\Phi\left(\theta, v_\gamma(x,t)\right) n_\gamma(x,t) ,\\
    \label{eq.Phi}
    \Phi(\theta,v) = \frac{1-v^2}{{2\left( 1 - v \cos{\theta}  \right)}^2} .
\end{align}
The mean velocity $\overline{v_\gamma}$ and the mean square velocity $\overline{v_\gamma^2}$ are
\begin{align}
    \label{eq.av_vx}
    \overline{v_\gamma}=\int_0^\pi \Phi(\theta, v) \cos{\theta} \sin{\theta}d\theta = \frac{1}{v} - \frac{1-v^2}{v^2}\text{ath}(v) , \\
    \overline{v_\gamma^2}=\int_0^\pi \Phi(\theta, v) \cos^2{\theta} \sin{\theta}d\theta = \frac{2\overline{v_\gamma}}{v} - 1 ,
\end{align}
where $\text{ath}(x)$ is the inverse hyperbolic tangent function. The results of the QED-PIC simulations show that Eq.~\eqref{eq.Phi} is a good approximation for the angular distribution of the \textit{involved} gamma-quanta [see Fig.~\ref{fig.ang} (a), (b)].

The $\chi$ parameter of the gamma-quanta propagating in the crossed electric and magnetic fields which is the case for both the plasma and the vacuum region can be calculated as follows
\begin{equation}
    \chi_\gamma = \frac{\varepsilon_\gamma \left| B - E\cos\theta \right|}{\mathcal{E}_S} = \frac{\varepsilon_\gamma E }{\mathcal{E}_S}\frac{1- v_x\cos\theta }{v_x},
\end{equation}
where we used~\eqref{eq.vdrift}.

Having specified the distribution function of the gamma-quanta completely it is possible to calculate sources $S[\alpha, pp]$ which correspond to the process of the pairs photoproduction
\begin{eqnarray}
    \begin{split}
        S[n,pp] =  n_\gamma\int\limits_0^\pi \Phi(\theta, v_\gamma) W_{pair}(\chi_\gamma,\varepsilon_\gamma)\times \\ 
        \times \sin\theta d\theta \equiv \overline{W_{pair}} n_\gamma ,
    \end{split}\\
    \begin{split}
        S[\varepsilon,pp] = \varepsilon_\gamma n_\gamma\int\limits_0^\pi \Phi(\theta, v_\gamma) W_{pair}(\chi_\gamma,\varepsilon_\gamma)\times \\ 
        \times \sin\theta d\theta \equiv \overline{W_{pair}} \varepsilon_\gamma n_\gamma ,
    \end{split}\\
    \begin{split}
        S[v,pp] = n_\gamma\int\limits_0^\pi \Phi(\theta, v_\gamma) W_{pair}(\chi_\gamma,\varepsilon_\gamma)\cos\theta \times \\ 
        \times \sin\theta d\theta \equiv \overline{V_{pair}} n_\gamma .
    \end{split}
\end{eqnarray}

\subsection{Pairs dynamics in the vacuum region}
Let us first examine the electrons and the positrons located in the vacuum region, where the number of particles is small so the collective plasma effects can be neglected. Thus the electromagnetic field in that region coinsides with the field of the incident radiation. In our case these electrons and positrons that are born in the vacuum region move in the field of a plane EM wave. Dynamics of a single particle in the plane wave is discussed in Appendix~\ref{app.Acceleration}. Computing the sources $S[\alpha, \beta]$ in the rhs of the~\eqref{eq.hd_p1}~--~\eqref{eq.hd_e1} requires knowledge of the particles distribution function. Although approximate particles trajectories can be found analytically, deriving explicit expression for the distribution function is practically unfeasible due to the fact that particles are being born at different time instances with varying initial conditions. But few observations will allow us to estimate $S[\alpha, \beta]$ using the different approach.
The first observation is that a strong EM plane wave pushes particles in the direction of its propagation, i.e. the $x$-axis. So after some time interval independently on its initial direction the particles momenta orient parallel to the $x$-axis, so that $v_x \approx 1$ [see Fig.~\ref{fig.scheme} (c)]. We neglect the duration of this momenta orientation, which allows us to approximate the fluxes of the particles density and particles energy density by simply multiplying these densities by the velocity $v_x \approx 1$.

The purpose of the continuity equations in the vacuum region essentially is to provide the values of the particles densities, energies and magnitude of the electric field at the plasma-vacuum interface (which we call occasionally the cascade front). So we are interested in the total contribution to these values by each particle during its motion from the moment of its birth in the vacuum region up to the moment of reaching the plasma boundary. We do this by defining the sources $S[\varepsilon,\beta]$ in the following way
\begin{equation}
    S[\varepsilon,\beta] = \int_0^\pi f_\gamma(x, \theta) W_{pair}(\chi_\gamma, \varepsilon_\gamma) \Delta\varepsilon_\beta \sin\theta d\theta ,
\end{equation}
where $\Delta\varepsilon_\beta$ is the total change of the energy due to the process $\beta$ of a single particle born at the point in space with the coordinate $x$ at the time instance $t$ while the particle stays in the vacuum region. 

The energy gained by the single particle in the plane wave can be evaluated as 
\begin{equation}
    \Delta\varepsilon_{acc} = \mu 2^{1/3} a_0^{2/3} \varepsilon_0^{1/3} \left( 1-\cos\theta \right)^{1/3} ,
\end{equation}
where $\varepsilon_0$ is the initial particle energy and $\mu$ is the parameter determining the time the particle spends in the vacuum region (see Appendix~\ref{app.Acceleration}). It depends on the time of the particle birth and the time when the particle crosses the plasma boundary. Finding the latter requires either building an independent model of the cascade front propagation or finding some heuristics both of which lays out of the scope of the paper. Instead we suppose that $\mu$ is constant during the whole cascade development and its value can be approximated by comparing the results of the model with the results of QED-PIC simulations, so $\mu$ is the first fitting parameter of our model. 
Noting that when pairs are born from gamma-quanta their average initial energy is approximately equal to half the energy of the parent gamma quanta $\varepsilon_\gamma$ we derive the expression for $S[\varepsilon,acc]$
\begin{equation}
    \label{eq.jE}
    \begin{split}
    S[\varepsilon, acc] =  E^{2/3} \varepsilon_\gamma^{1/3} \mu n_\gamma \int_0^\pi \Phi(\theta, v_\gamma) W_{pair}(\chi_\gamma, \varepsilon_\gamma) \times \\
    \times {\left( 1-\cos\theta \right)}^{1/3}\sin\theta d\theta \equiv E^{2/3} \varepsilon_\gamma^{1/3} \mu \overline{G_{rad}} n_\gamma.
    \end{split}
\end{equation}

The total energy loss due to photon emission of a single particle can be calculated as follows (see Appendix~\ref{app.Acceleration})
\begin{eqnarray}
    \label{eq.GammaRad0}
    \Delta\varepsilon_{rad}=\int\limits_0^{\sqrt{4\mu^3/9}} I_{rad}(\chi)dt , \\
    \frac{d\chi}{dt}=-\frac{\chi_0}{\varepsilon_0+{\left[ \frac{9}{2} E^2 \varepsilon_0 t^2 (1-\cos\theta)  \right]}^{1/3}}I_{rad}(\chi), \\
    \chi_0 = \frac{\chi_\gamma}{2},\ \varepsilon_0 = \frac{\varepsilon_\gamma}{2}.
\end{eqnarray}
So the source term $S[\varepsilon_p, rad]$ takes the form
\begin{equation}
    \begin{split}
        S[\varepsilon_p, rad] = n_\gamma \int_0^\pi \Phi(\theta, v_\gamma) W_{pair}(\chi_\gamma, \varepsilon_\gamma) \times \\ 
        \times \Delta\varepsilon_{rad}\sin\theta d\theta \equiv \overline{I_{vac}} n_\gamma .
    \end{split}
\end{equation}

\subsection{Pairs dynamics in the plasma region}
\label{sub.Pairs}

\begin{figure}
    \includegraphics[width=60mm]{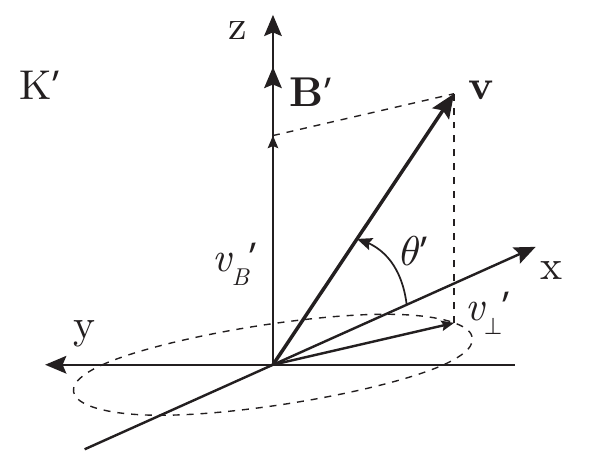}
    \caption{\label{fig.plasma} Relation between the velocity vector and the magnetic field vector in the reference frame $K'$ moving along the $x$-axis with the velocity $v_x=E/B$.}
\end{figure}

In the plasma region at each point in space there exists an instantaneous reference frame $K'$ moving with the velocity $v_x(x,t)\approx E/B$ in which only magnetic field presents. It is convenient to obtain some results in that reference frame. In $K'$ frame electrons and positrons move along the magnetic field with the velocity $v_B'$ and revolve in the plane perpendicular to it with the velocity $v_\perp'$ (see Fig.~\ref{fig.plasma}). We will assume that the particles remain ultrarelativistic in that reference frame, so ${v_\perp'}^2+{v_B'}^2\approx1$. Moving along the magnetic field results in non-zero average current which has to be accounted in the Maxwell's equations, while revolving in the magnetic field gives zero average current but it is responsible for producing gamma-quanta. In particular the Lorentz-invariant QED parameter $\chi$ for pairs can be written as
\begin{equation}
    \chi_p=\frac{v_\perp'\varepsilon_p'B'}{E_S} .
\end{equation}
We express the primed values through the values in the laboratory reference frame as follows: $B'=B\sqrt{1-(E/B)^2}$, $\varepsilon_p'=\varepsilon_p\sqrt{1-(E/B)^2}$, where we used the fact that the average particle momentum along the $x$-axis is equal to $\gamma v_x$ and $v_x=E/B$. The final form of the expression for $\chi$ is
\begin{equation}
    \label{eq.chip}
    \chi_e = \frac{v_\perp'\varepsilon_p E}{E_S} \frac{1-v_x^2}{v_x} .
\end{equation}
Due to particles revolving around the direction of the magnetic field we can assume that the angular distribution of the particles in the $K'$ reference frame is close to uniform. Applying Lorentz transformation we derive the following distribution function of pairs in the laboratory reference frame
\begin{align}
    \label{eq.distr}
    f_p(t,x,\theta)=\Phi\left(\theta, v_x(x,t)\right) n_p(x,t) .
\end{align}
where $\Phi$ is defined the same as in Eq.~\eqref{eq.Phi}
\begin{equation}
    \Phi(\theta,v) = \frac{1-v^2}{{2\left( 1 - v \cos{\theta}  \right)}^2} . \nonumber
\end{equation}
Same as for the gamma-quanta the results of the QED-PIC simulations demonstrate that Eq.~\eqref{eq.distr} is a good approximation for the angular distribution of the pairs [see Fig.~\ref{fig.ang} (c), (d)]. Below we will show that the velocity $v_x$ can be calculated from the local values of the electric field and plasma density. That is why we do not include the continuity equation for the density of the longitudinal velocity of the pairs similar to Eq.~\eqref{eq.hd_g2}. Also in the case of the pairs we choose to neglect the difference between the velocity $v$ of the reference frame in which the particles angular distribution is uniform and the actual mean velocity $\overline{v}$ calculated from that distribution, the maximum difference between which is less then $0.2$, according to Eq.~\eqref{eq.av_vx}.

Because $\chi_p$ does not depend on $\theta$ we calculate the sources $S[\alpha, rad_c]$ as follows
\begin{eqnarray}
    \begin{split}
        S[n, rad_c] = n_p \int\limits_0^\pi \Phi(\theta, v_x) W_{rad}(\chi_p, \varepsilon_p) \times \\
        \times \sin\theta d\theta = W_{rad}(\chi_p, \varepsilon_p) n_p \equiv \overline{W_{pl}} n_p,
    \end{split} \\
    \begin{split}
        S[\varepsilon, rad_c] =  n_p \int\limits_0^\pi \Phi(\theta, v_x) I_{rad}(\chi_p) \times \\
        \times \sin\theta d\theta = I_{rad}(\chi_p) n_p \equiv \overline{I_{pl}} n_p,
    \end{split}\\
    \begin{split}
        S[n, rad_c] =  n_p \int\limits_0^\pi \Phi(\theta, v_x) \cos\theta W_{rad}(\chi_p, \varepsilon_p) \times \\
        \times \sin\theta d\theta =  W_{rad}(\chi_p,\gamma) v_x n_p  \equiv \overline{W_{pl}} v_x n_p.
    \end{split}
\end{eqnarray}

The total current density of the particles averaged over the characteristic period of the Larmor oscillations $\tau_B = \varepsilon_p/B$ is
\begin{align}
    \label{eq.j}
    \mathbf{j} = 2 n_p \frac{\mathbf B}{B} v_B \sqrt{1-v_x^2}, \\
    v_B = v_B' \frac{2}{\pi}\frac{\arccos{(v_x\sqrt{1-v_x^2})}}{\sqrt{1-v_x^2(1-v_x^2)}}\sim v_B'.
\end{align}
The factor $2$ is from the fact that currents of the electrons and the positrons are co-directional. This comes from the observation that in the laboratory reference frame the electric and the magnetic fields are actually not exactly perpendicular. It means that in the reference frame $K'$ there exist a small electric field directed along or opposite to the magnetic one (depending on the sign of the $\mathbf{E\cdot B}$ product). Presence of that field leads to the average electrons velocity being counter-directed to it and the average positrons velocity being co-directed to it. This is not the case for the longitudinal motion of the pairs which does not depend on the charge sign, thus currents of the electrons and the positrons eliminate each other along the $x$-axis but sum up in the $yz$-plane. This fact also hints that the electron-positron plasma is actually a conducting media thus some electromagnetic energy absorption also occurs in that region, though it is significantly smaller than the absorption in the vacuum region which is seen in the QED-PIC simulations [see Fig.~\ref{fig.assumptions} (a)] and thus we do not account it in our model.

 The value of $v_B$ averaged over the particles which we denote as $\nu$ is the second fitting parameter of our model. We can roughly estimate it by noting that for a single particle the value of $v_B'$ cannot exceed its initial value during the particle motion. The particles enter the plasma region after being accelerated by the laser pulse with predominantly longitudinal velocity, i.e. velocity along the $x$-axis, so the initial projection of the particles velocity onto the magnetic field which lies in the $yz$-plane is small. So we expect our model to give valid results with the values of $\nu$ closer to zero rather than unity.

 The electrodynamical properties of the medium which response to the plane EM wave consists of inducing the current along the magnetic field is examined in Appendix~\ref{app.Electrodynamics}. The main conclusion is that the ratio between the electric and the magnetic field in such media can be expressed via its density and the electric field amplitude
 \begin{equation}
    \label{eq.vx}
     \frac{E}{B}=v_x=\sqrt{\frac{2}{1+\sqrt{1+\left(4n_p\nu/E \right)^2}}} .
 \end{equation}
 Validity of Eq.~\eqref{eq.vx} is verified by direct comaprison against the mean values of the longitudinal velocity of the particles computed in the PIC simulation as show in Fig.~\ref{fig.assumptions} (b).

\section{Model formulation and comparison with QED-PIC simulations}
\label{sec.Discussion}
The last undefined terms are the $\psi_{vac}$ and $\psi_{pl}$ which determine the vacuum and the plasma regions, respectively, in space. We note that the longitudinal velocity $v_x$ defined in Eq.~\eqref{eq.vx} actually demarcates these regions: in the vacuum region $v_x\approx 1$ and in the plasma region $v_x < 1$. So we choose $\psi_{vac}$ and $\psi_{pl}$ in the following way
\begin{align}
    \psi_{vac} = v_x^M \\
    \psi_{pl} =  1 - v_x^M
\end{align}
where $M\sim 10$ is a constant. We choose the exact value of this parameter by defining the upper threshold for $v_x$ above which we assume plasma to be rarefied enough not to cause any collective effects. So we choose this threshold value to be $0.7$ and $M=8$.

The final set of the cascade model equations is the following
\begin{widetext}
\begin{align}
    \label{eq.fin1}
    \frac{\partial }{\partial t}n_p  +\frac{\partial}{\partial x}\left( v_x n_p \right) &= \overline{W_{pair}} n_\gamma  ,\\
    \label{eq.fin2}
    \frac{\partial}{\partial t}\left( \varepsilon_p n_p \right)  + \frac{\partial}{\partial x}\left( v_p \varepsilon_p n_p \right)  &= \overline{W_{pair}} n_\gamma \frac{\varepsilon_\gamma}{2}  +\left( \mu E^{2/3}\varepsilon_\gamma^{1/3}\overline{G_{rad}} - \overline{I_{vac}} \right)n_\gamma \psi_{vac} - \overline{I_{pl}} n_p \psi_{pl} ,\\
    \label{eq.fin3}
    \frac{\partial }{\partial t}n_\gamma  +\frac{\partial}{\partial x}\left( \overline{v_\gamma} n_\gamma \right)&=-\overline{W_{pair}} n_\gamma + 2\overline{W_{rad}} n_p \psi_{pl} ,\\
    \label{eq.fin4}
    \frac{\partial}{\partial t}\left( \overline{v_\gamma} n_\gamma \right) + \frac{\partial}{\partial x}\left( \overline{v_\gamma^2} n_\gamma \right) &= -\overline{V_{pair}} n_\gamma + 2 \overline{V_{rad}} n_p \psi_{pl} ,\\
    \label{eq.fin5}
    \frac{\partial}{\partial t}\left( \varepsilon_\gamma n_\gamma \right) +\frac{\partial}{\partial x}\left( \overline{v_\gamma} \varepsilon_\gamma n_\gamma \right) &=-\overline{W_{pair}}n_\gamma\varepsilon_\gamma   +2 \overline{I_{pl}} n_p \psi_{pl} ,\\
    \label{eq.fin6}
    \frac{\partial }{\partial t}\left( \frac{E^2 + E^2/v_x^2}{2} \right) +\frac{\partial}{\partial x} \left( \frac{E^2}{v_x} \right) &=  -2\mu E^{2/3}\varepsilon_\gamma^{1/3}\overline{G_{rad}} n_\gamma\psi_{vac}  ,\\
    \label{eq.fin7}
    \frac{\partial }{\partial t}\Sigma_\gamma &=  \overline{I_{vac}} n_\gamma \psi_{vac} .
\end{align}
\end{widetext}
It is important to note that the energy is conserved in the model, i.e.
\begin{equation}
    \int \left( 2n_p\varepsilon_p + n_\gamma \varepsilon_\gamma + \frac{E^2 + B^2}{2} \right) dx + \Sigma_\gamma = \text{const} .
\end{equation}

\begin{figure*}
    \includegraphics[width=175mm]{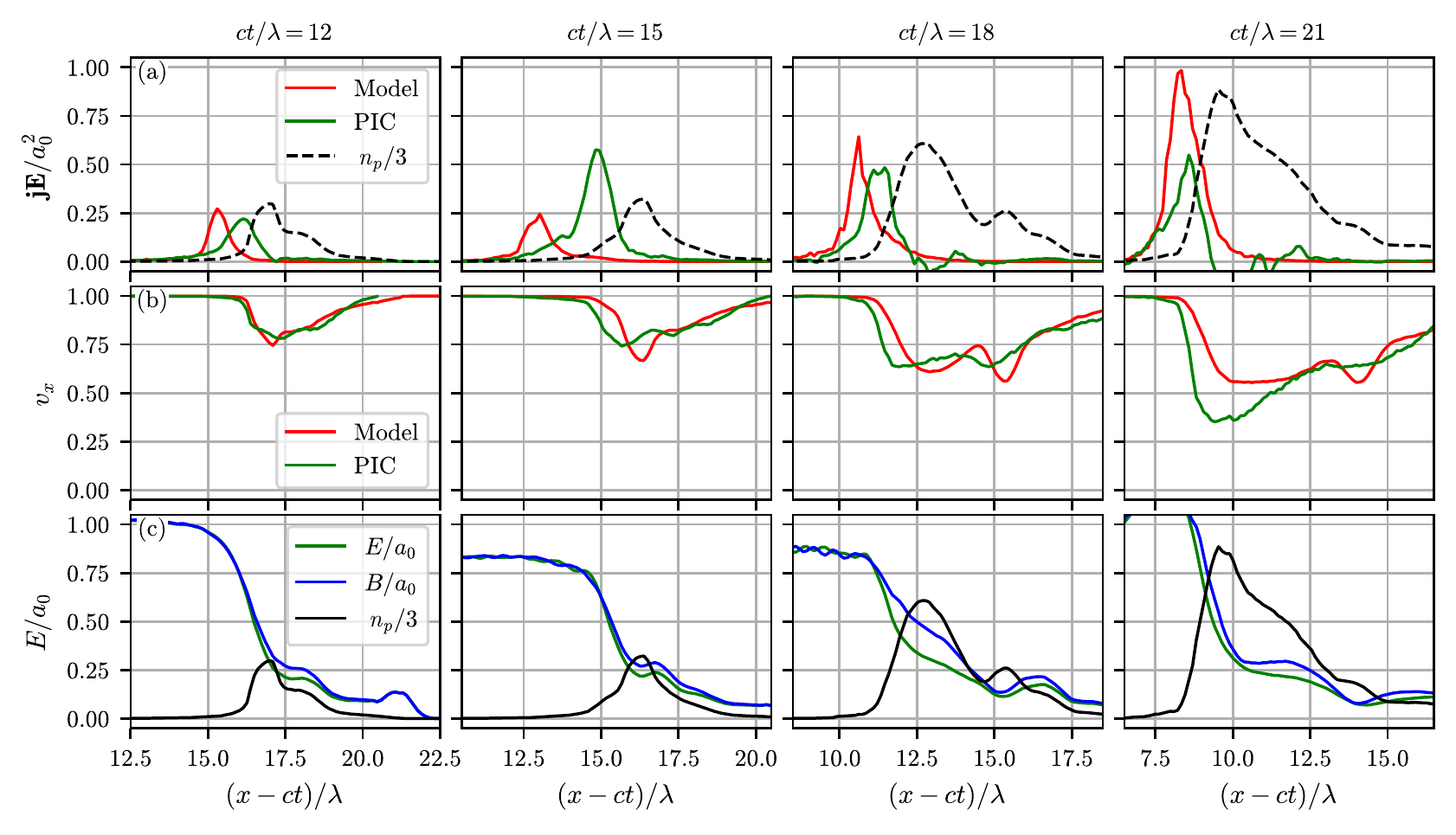}
    \caption{\label{fig.assumptions} Validation of the approximations used in the model. (a) The value of the $\mathbf{jE}$ product computed from Eq.~\eqref{eq.jE} and computed directly from the results of the QED-PIC simulations (red lines). Note that the absorption is large in the vacuum region where pairs density is small (dashed black line) and is almost negligible in the dense plasma region. Also note that approximate value is shifted along the $x$-axis; it corresponds to the fact that we compute $\mathbf{jE}$ for each particle at the moment of its birth, while actually $\mathbf{jE}$ is gained during all the time the particle spends on the vacuum region. (b) The mean longitudinal velocity of the pairs computed from Eq.~\eqref{eq.vx} and computed directly from the results of the QED-PIC simulations (red lines). (c) The distribution of the electric field (green line), magnetic field (blue line) and the plasma density (black line). Each plot is computed for 2~--~$\lambda$ vicinity of the center of the laser pulse.}
\end{figure*}

The main approximations~---~Eqs.~\eqref{eq.jE} and ~\eqref{eq.vx}~---~are directly validated based on the results of the QED-PIC simulations (see Fig.~\ref{fig.assumptions}). PIC simulations were performed using the QUILL code~\cite{QUILL}, which enables modelling of the QED effects via the Monte-Carlo method. The initial distrubution of the EM fields has a form of the plane wave with the wavelength $\lambda=2\pi c/\omega_{L}=1 \text{ }\mu \text{m}$ and the amplitude $a_0$ propagating along the $x$-axis with the spatio-temporal envelope given by the following expression
\begin{equation}
a(x,y,z) =  \cos^2 \left( \frac{ \pi }{2}   \frac{x^4}{\sigma_x^4 } \right) \cos^2 \left( \frac{ \pi}{2}   \frac{\left( y^2 + z^2 \right)^2}{\sigma_r^4 } \right)
\end{equation}
The transverse spatial size of the laser pulse is $2\sigma_r = 18~\mu$m and the pulse duration is $60.5$~fs ($2 \sigma_x = 18.15~\mu$m). The simulation box size is $30\lambda\times 30 \lambda \times 30 \lambda$, the grid size is $3000 \times 300 \times 300$. 
As discussed in Ref.~\cite{Samsonov2019} the final stage of the QED cascade development in a single laser pulse almost does not depend on the seed, so we choose the seed in the form of a short gamma-bunch counter-propagating to the laser pulse in order not to introduce the electron-ion plasma to the interaction which is significantly different from the forming electron-positron plasma. The initial seed of that form can be used in our model by initializing $v_\gamma(t=0)\approx-1$. The density distributions in our model and PIC simulation coinside and are expressed by the following formula: $n_\gamma(t=0)=n_0\max\left\{0, 1 - (x-x_0)^2/w_\gamma^2  \right\}$, where $w_\gamma$ is the half-width of the bunch and $x_0$ is the position of its center.

The Eqs.\eqref{eq.fin1}--\eqref{eq.fin6} are solved numerically using the method of lines (MOL): partial derivatives $\partial/\partial x$ are approximated with the finite differences to derive the system of the ODEs which is solved using the explicit Runge-Kutta method. This scheme is not internally conservative so the energy conservation is done manually at each integration step by clipping the derivative $\partial n_\gamma /\partial t$ so that the total energy does not grow. The relative error gained from that procedure turns out to be acceptably small.

\begin{figure*}
    \includegraphics[width=170mm]{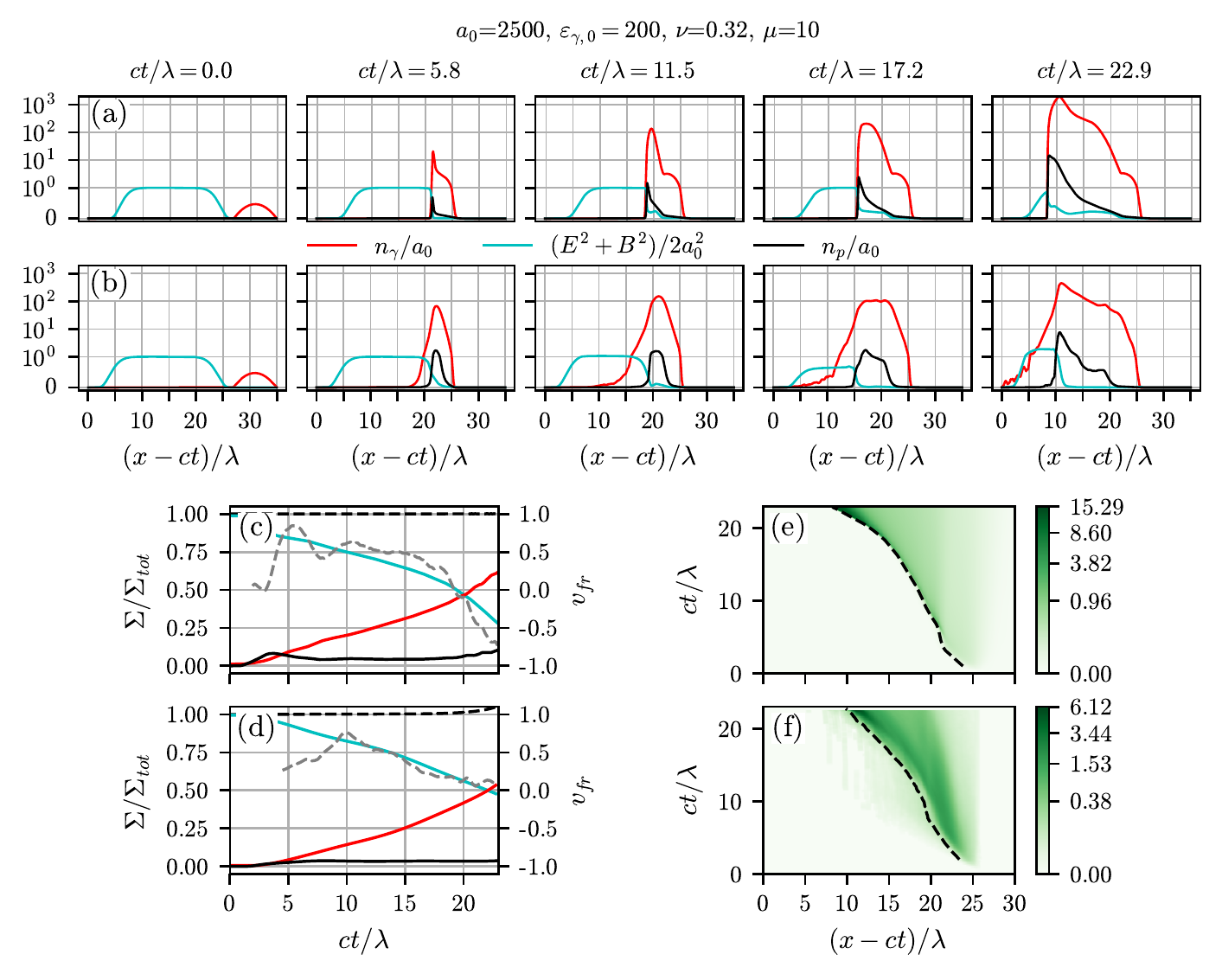}
    \caption{\label{fig.sol2500} Comparison between the solution of Eqs.~\eqref{eq.fin1}--\eqref{eq.fin6} and the results of the QED-PIC simulations with identical initial conditions in the form of the laser pulse with the amplitude $a_0=2500$ and the counter-propagating beam of gamma-quanta with the energy $200 m_e c^2$ and the density $0.5 n_{cr}$ ($n_{cr}$ is the critical density for the laser wavelength of 1 $\mu$m). The distributions of the gamma-quanta density (red line), the EM energy density (cyan line) and the plasma density (black line) at different time instances observed in the model solution (a) and in the PIC simulation (b). Note that the scale of the vertical axis is linear in the range $[0,1]$ and logarithmic in the range $[1,+\infty]$.
    The energy balance: the total energy of the pairs (solid black line), the gamma-quanta (red line) and the electromagnetic energy (cyan line) normalized to the initial total energy of the system (dashed black line); the velocity of the plasma boundary (grey dashed line) observed in the model solution (c) and in the PIC simulation (d).
    The distribution of the pairs in the $x-t$ plane (color map) and the position of the plasma boundary (black dashed line) in the model solution (e) and in the PIC simulation (f).
    The numeric values of the fitting parameters used in the model solution are: $\nu=0.3$, $\mu=0.4$.}
\end{figure*}

\begin{figure*}
    \includegraphics[width=170mm]{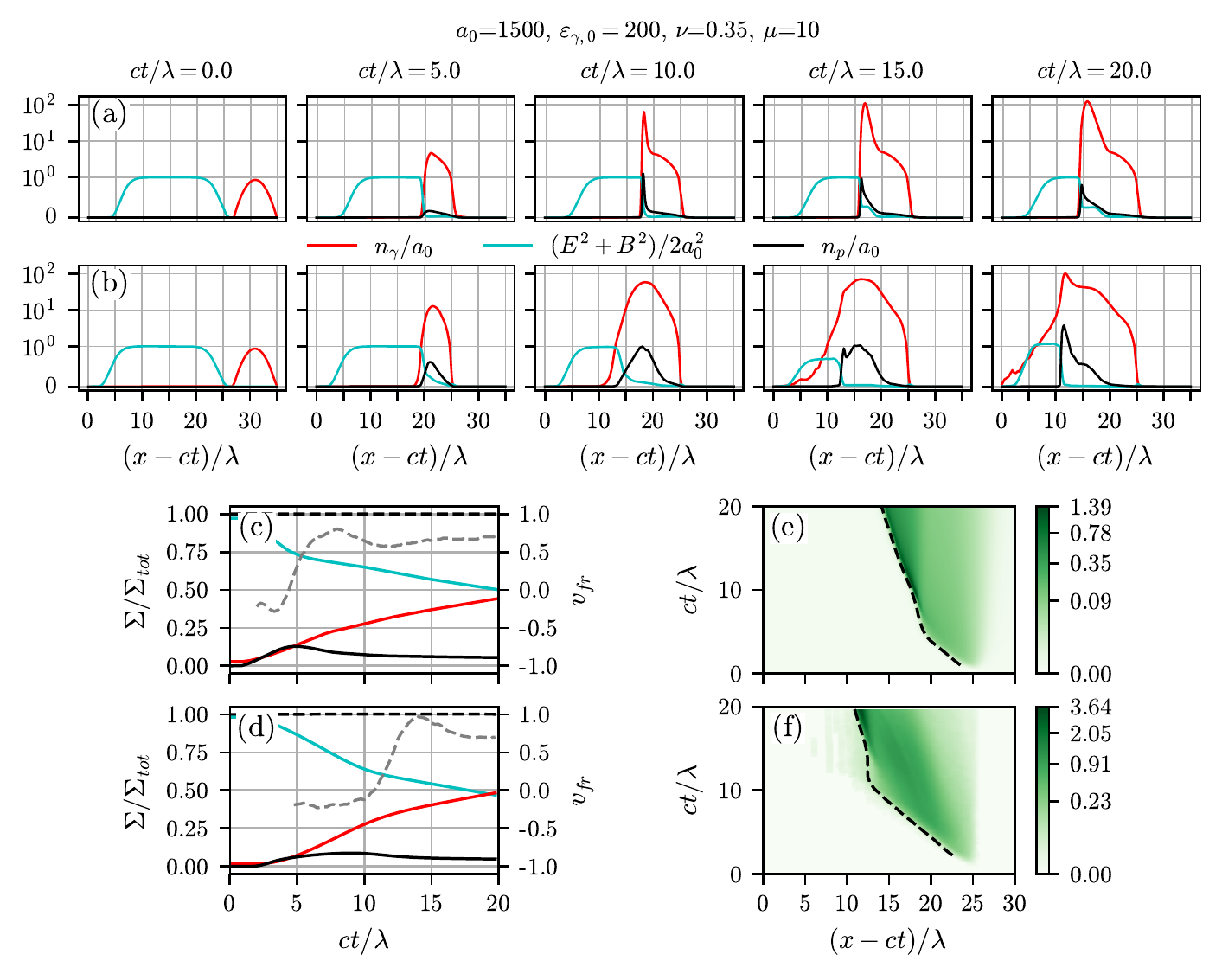}
    \caption{\label{fig.sol1500} Same as in Fig.~\ref{fig.sol2500} but for $a_0=1500$, $n_{\gamma,0}=n_{cr}$. The values of the fitting parameters are: $\nu=0.3$, $\mu=0.4$.}
\end{figure*}

\begin{figure*}
    \includegraphics[width=170mm]{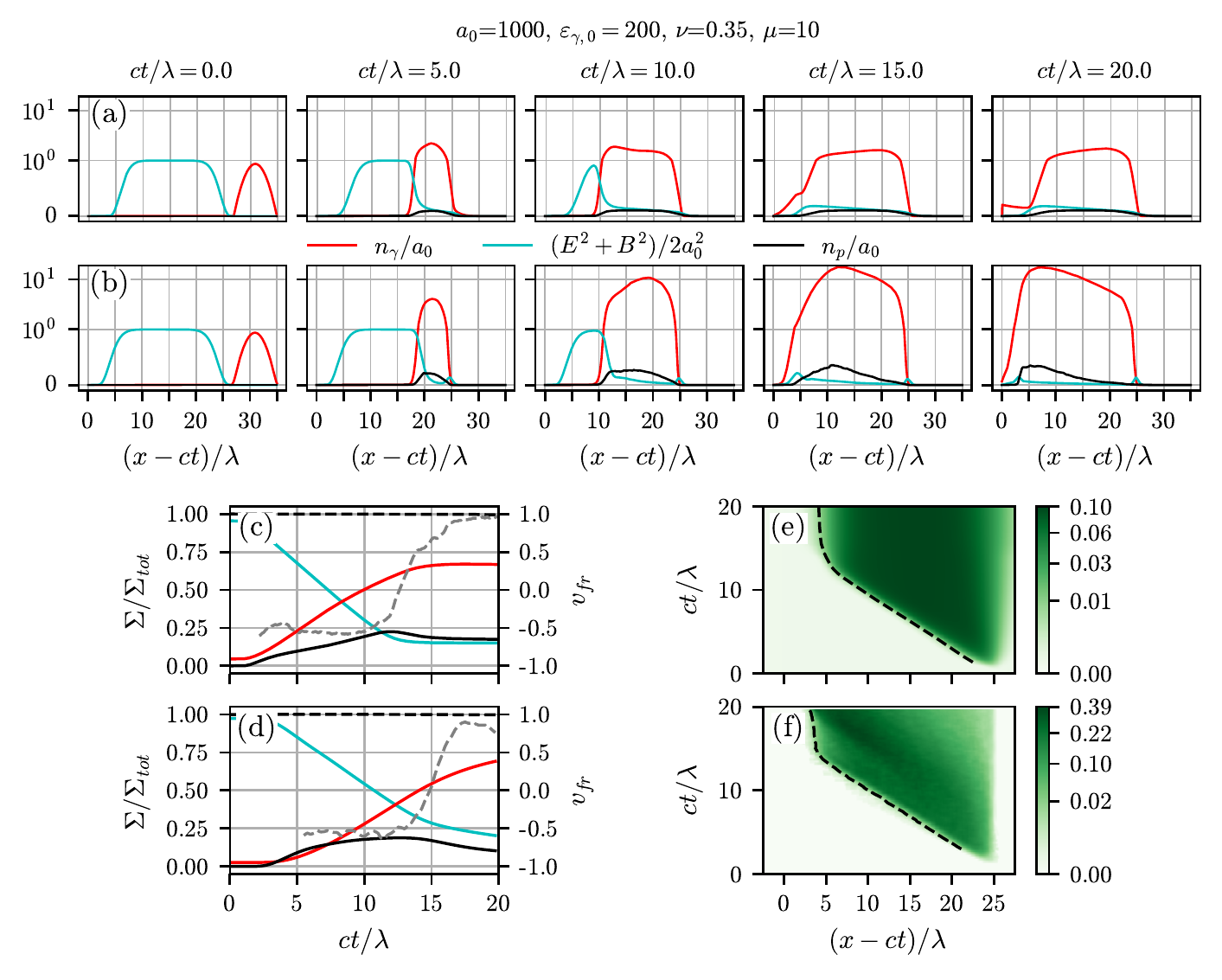}
    \caption{\label{fig.sol1000} Same as in Fig.~\ref{fig.sol2500} but for $a_0=1000$, $n_{\gamma,0}=n_{cr}$. The values of the fitting parameters are: $\nu=0.25$, $\mu=4$.}
\end{figure*}

Direct comparisons between the solutions of Eqs.~\eqref{eq.fin1}~--~\eqref{eq.fin6} and the results of the QED-PIC simulations are shown in Figs.~\ref{fig.sol2500}~--~\ref{fig.sol1000}. Our model coinsides with the results of the full QED-PIC simulations qualitatively good in terms of the distributions of the particles and the electromagnetic field as well as the energy balance. Also we can clearly see the different regimes of the cascade development in both cases. 

The first regime is observed when $a_0$ of the laser pulse is not big enough or the gamma-bunch is not dense enough. In that case the density of the produced electron-positron plasma does not reach the relativistic critical density so that $v_x\approx 1$, i.e. the collective plasma effects do not occur. In this case the plasma region is not present at all and the newly born particles move in the unaltered field of the laser pulse which is close to the plane wave. As discussed in~\cite{DiPiazza2012, Bulanov13a, narozhny2015quantum, mironov2017observable} in that case $\chi$ of the pairs does not grow during its motion in the plane wave. But after each act of the gamma-quant emission it splits between the parent and the child particles so after few generations $\chi$ of all the particles becomes negligibly small so the cascading ceases. Thus for small enough $a_0$ the gamma-quanta of the gamma-bunch decay into pairs levaing the `trail' of the electrons and positrons which are accelerated forwards and co-propagate with the laser pulse. Although the density of the plasma is small the total number of the pairs can be big enough so that the significant portion of the laser energy is transferred to them [see Fig.~\ref{fig.sol1000} (c), (d)]. Because in this regime all the particles propagate independently to each other the cascade front propagates with almost constant velocity $v_{fr}\approx -0.5$.

In the second regime the cascade develops as discussed in Sec.~\ref{sec.Introduction}. The peak of the pairs density propagates towards the laser with a much slower velocity (relative to the leading edge of the laser pulse) than in the first regime. Moreover the density of the plasma grows in time in contrast to the first regime where the plasma density at each point stays almost the same after the initial gamma-bunch passes that point. As mentioned in Sec.~\ref{sub.Pairs} the dense electron-positron plasma actually almost does not absorb the laser field that is why despite the fact that in this regime the total number of the pairs is much larger than in the first regime, the rates of the energy transfer from the EM field to the pairs are close to each other in both regimes.

If the $a_0$ lays in-between the values of $a_0$ at which either the first or the second regime is observed, at the initial stage the cascade resembles the S-type cascade which is clearly indicated by the negative value of the velocity of the cascade front [see grey dashed line in Fig.~\ref{fig.sol1500} (c), (d)]. At some point the density of the pairs becomes large enough to alter the laser propagation and to shift the cascade dynamics to the self-sustained regime. The change between these two regimes is indicated by abrupt change in the velocity of the cascade front. The initial stage (stage of the S-type cascade) can also be seen for larger values of $a_0$ (see Fig.~\ref{fig.sol2500}), though it is much shorter and is hardly pronounced in the results of the QED-PIC simulations.

There are some features that are not captured by our model which are worth noting. Firstly, in the PIC simulation there is a distinct tail of the gamma-quanta spatial distribution counter-propagating to the laser pulse. These gamma-quanta have relatively low energy and thus are unable to photoproduce pairs. Our model predicts that the edge of the plasma and gamma-quanta distributions almost competely coincide. The total energy carried away by this sort of gamma-quanta is insignificant so this feature is not crucial for the cascade development. The reason that our model cannot capture this feature is the fact that we assume the distribution functions to be monoenergetic. Higher accuracy can be obtained if we would split the gamma-quanta into several groups with different energies and describe them separately; then this feature would be present in our solutions. But as already mentioned in Sec.~\ref{sec.Main} it will greatly complicate the model but will not lead to significant qualitative changes in the solutions.
Secondly, both the total number of the pairs and peak plasma density are larger in the QED-PIC simulation than in our model solution for smaller values of $a_0$. One of the reasons behind these discrepancies is that our model is 1-dimensional and thus does not describe the laser pulse diffraction. In the 3D-PIC simulations the simulation box is always limited thus the pure plane wave cannot be achieved in the simulations. It leads to the fact that the envelope of the laser pulse evolves so that the magnitude of laser pulse in the vacuum region may differ from its initial value $a_0$ [see Fig~\ref{fig.assumptions} (c)]. The increase of the effective laser pulse intensity generally leads to the increase of the probabilities of the QED processes and thus more abundant pairs photoproduction. The net effect of the inconstancy of the laser pulse intensity can be partially accounted by choosing the value of $a_0$ in the model solution larger than the value set initially in the PIC simulation. 

\section{Conclusion}
\label{sec.Conclusion}
We have developed a self-consistent model of the QED cascade development in an extremely intense laser pulse. The complete description of that interaction requires solving the Maxwell's equations along with the kinetic equations for electrons, positrons and gamma-photons. This system of equations is too complex for analytical methods and is usually solved numerically with QED-PIC codes consuming a lot of computational resources. In order to derive the reduced equations for the computationally light model we adopt some assumptions. The main of them are: quasi-one-dimensional hydrodynamics; locally quasi-monoenergetic distribution function for the particles; the plane wave approximation for the laser radiation. The derived simplified system of the equations is written in the closed form and is solved numerically. Despite the complexity and non-linearity of the cascade dynamics it turned out that a relatively simple one-dimensional model can predict its development qualitatively, e.g.~the macroscopic spatio-temporal particles distribution as well as the energy balance in the system in our model and in the results of QED-PIC simulations coincide approximately. These facts serve as the justification of the analytical reasoning behind the model and hence our understanding of the phenomenon. Although there are several discrepancies between the predictions of our model and results of the QED-PIC simulations, the reasons behind them are identified and for some the method to resolve them is discussed.

We stress that the model is suitable for the complete description of the QED cascade development in the single laser pulse.  For example various regimes of the interaction based on the intensity of the laser pulse are revealed in both full QED-PIC simulations and our model.

We believe that the model also can be adapted to explore regimes of QED cascades in different environments: laser interaction with the targets including foils, particle beams or gamma-quanta, beam-beam interaction etc. Therefore further extensive test of the developed model is planned to be carried out in future.

\begin{acknowledgments}
    This research was supported by: Russian Science Foundation (Grant No.~20-12-00077), Foundation for the advancement of theoretical physics and mathematics “BASIS” (Grant No.~19-1-5-10-1).
\end{acknowledgments}

\appendix

\section{Particle acceleration in the plane wave}
\label{app.Acceleration}
Let us consider a single electron motion in the field of a plane circularly polarized electromagnetic wave. The vector potential of the wave is chosen as follows
\begin{align}
    \mathbf{A} = a_0 \left( \mathbf{e_y} \cos\xi + \mathbf{e_z} \sin\xi \right) ,\\
    \xi = t - x .
\end{align}
The Hamiltonian for the problem is
\begin{equation}
    \label{eq.A.Ham}
    H = \sqrt{1 + {\left( \mathbfcal{P} + \mathbf{A} \right)}^2},
\end{equation}
where $\mathbfcal{P} = \mathbf{p - A}$ is the generalized momentum of the electron with the momentum $\mathbf{p}$.
Deriving the motion equations we get
\begin{align}
    \frac{d \mathcal{P}_y}{dt} = -\frac{\partial H}{\partial y} = 0 ,\\
    \frac{d \mathcal{P}_z}{dt} = -\frac{\partial H}{\partial z} = 0 ,\\
    \frac{d \mathcal{P}_x}{dt} = \frac{d p_x}{dt} = -\frac{\partial H}{\partial x} = \frac{\partial H}{\partial\xi} ,\\
    \frac{d \varepsilon}{dt} = \frac{d H}{dt} = \frac{\partial H}{\partial t} +\cancelto{0}{\left[ H\mathcal{P} \right]} = \frac{\partial H}{\partial\xi}.
\end{align}
From these Eqs. we derive that
\begin{align}
    \label{eq.A.py}
    p_y + A_y = \text{const} ,\\
    \label{eq.A.pz}
    p_z + A_z = \text{const} ,\\
    \label{eq.A.px}
    p_x - \varepsilon = \text{const}.
\end{align}
Let us define the initial conditions of the electron
\begin{align}
    p_x(t=t_0) = p_0 \cos\theta ,\\
    p_y(t=t_0) = p_0 \sin\theta ,\\
    p_z(t=t_0) = 0 ,\\
    x(t=t_0) = x_0 ,\\
    \xi(t=t_0) \equiv t_0 - x_0 ,\\
    \varepsilon_0\equiv\sqrt{1+p_0^2}.
\end{align}
We can then rewrite Eqs.~\eqref{eq.A.py}~--~\eqref{eq.A.px}
\begin{align}
    p_y = p_0\sin\theta+a_0\left( \cos\xi_0 - \cos\xi \right) ,\\
    p_z = a_0\left( \sin\xi_0 - \sin\xi \right) ,\\
    \label{eq.a.g1}
    \varepsilon = p_x + \rho_0, \\
    \rho_0 = \varepsilon_0-p_0\cos\theta.
\end{align}
The Eq.~\eqref{eq.A.Ham} can be rewritten in the following form
\begin{equation}
    \label{eq.a.g2}
    \varepsilon = \sqrt{1 + p_x^2 + p_y^2 + p_z^2} .
\end{equation}
Combining Eqs.~\eqref{eq.a.g1}--~\eqref{eq.a.g2} we get the equation for $\varepsilon$
\begin{equation}
    \begin{split}
        \varepsilon =  \varepsilon^2 - \left\{ 1 - a_0^2 {\left( \sin\xi - \sin\xi_0 \right)}^2  - \right. \\
        - \left. {\left[ p_0\sin\theta + a_0\left( \sin\xi - \sin\xi_0 \right) \right]}^2 \right\}^{1/2} + \rho_0.
    \end{split}
\end{equation}
The energy gain $\Delta\varepsilon=\varepsilon-\varepsilon_0$ can be analytically found
\begin{equation}
    \label{eq.a.dg}
    \Delta\varepsilon = \frac{2a_0p_0}{\rho_0}\sin\frac{\Delta\xi}{2}\left( \frac{a_0}{p_0}\sin\frac{\Delta\xi}{2} - \sin\theta\sin\frac{\xi+\xi_0}{2} \right) ,
\end{equation}
where $\xi$ is the current phase of the electron. If we assume that $1\ll p_0\ll a_0$ then the second term in the brackets of Eq.~\eqref{eq.a.dg} is insignificant for any $\xi$ and $\xi_0$ compared to the first term and so we can omit it. Another reasoning that allows us not to account the second term is the fact that it depends on the abolute phase in which the particle is located and thus its value averaged over the wave period is equal to zero.
\begin{equation}
    \label{eq.A.G}
    \Delta\varepsilon\approx\frac{2a_0^2}{\rho_0}\sin^2\left( \frac{\Delta\xi}{2} \right).
\end{equation}

Next we calculate $\Delta \xi$
\begin{equation}
    \begin{split}
    \frac{d\xi}{dx}=\frac{d}{dx}\left( t - x \right) = \frac{1}{v_x} - 1 = \\ 
    = \frac{\varepsilon}{p_x}-1=\frac{\varepsilon-p_x}{p_x}=\frac{\rho_0}{p_x},
    \end{split}
\end{equation}
\begin{equation}
    \label{eq.A.deltax}
    \begin{split}
        \Delta x \rho_0 = \int\limits_{\xi_0}^{\xi_0+\Delta\xi} p_x d\xi = \int\limits_{\xi_0}^{\xi_0+\Delta\xi} \left( \Delta\varepsilon+p_0\cos\theta \right) d\xi = \\
        = \frac{a_0^2}{\rho_0}\left( \Delta\xi-\sin\Delta\xi \right) + p_0\cos\theta\Delta\xi.
    \end{split}
\end{equation}
If we assume again that $a_0 \gg p_0$ then Eq.~\eqref{eq.A.deltax} can be rewritten as follows
\begin{equation}
    \label{eq.A.deltaxi}
    \Delta\xi - \sin\Delta\xi = \frac{\rho_0^2}{a_0^2}\Delta x.
\end{equation}
Note that $\rho_0 < p_0 \ll a_0$ and $\Delta x$ is the distance along the $x$-axis between the initial and final position of the electron. In the case of our cascade model the initial position coinsides with the position of the parent gamma-quant decay and the final position is the position of the cascade front. QED-PIC simulations show that the distance $\Delta x$ does not exceed several $\lambda$ or in dimensionless variables it means that $\Delta x/2\pi \sim 1$. So in Eq.~\eqref{eq.A.deltaxi} we can assume that $\Delta\xi\ll1$ and leave only the first non-vanishing term in the lhs expansion
\begin{align}
    \frac{\Delta\xi^3}{6}\approx\frac{\rho_0^2}{a_0^2}\Delta x, \\
    \Delta\xi \approx {\left( 6\Delta x\frac{\rho_0^2}{a_0^2} \right)}^{1/3}.
\end{align}
Substituting this solution to Eq.~\eqref{eq.A.G} we get
\begin{equation}
    \label{eq.A.DG}
    \Delta\varepsilon\equiv\Delta\varepsilon_{acc}\approx {\left[ \frac{9}{2} a_0^2 \varepsilon_0 {\Delta x}^2 (1-\cos\theta)  \right]}^{1/3},
\end{equation}
where we also assumed that $\rho_0\approx \varepsilon_0\left( 1-\cos\theta \right)$
The duration of the particle motion can be also calculated
\begin{equation}
    \Delta\xi = \Delta t - \Delta x,
\end{equation}
and under assumption that $\Delta \xi \ll \Delta x$ which follows from the assumption $p_0 \ll a_0$ we get that $\Delta x \approx \Delta t$. It is an obvious conclusion because under our assumptions the particle can be considered ultrarelativistic and its velocity along the $x$-axis is close to the speed of light.

Below we estimate the radiative losses during the particle motion in the plane wave. The governing parameter $\chi$ in the plane wave is caluclated as follows
\begin{equation}
    \label{eq.A.chi}
    \chi = \frac{\varepsilon}{\mathcal{E}_S} \sqrt{ \left(\mathbf{E}+\mathbf{v}\times\mathbf{B}\right)^{2}-\left(\mathbf{v}\cdot\mathbf{E}\right)^{2} } = \frac{a_0}{\mathcal{E}_S} \left( \varepsilon - p_x \right) .
\end{equation}
Where $\mathbf{E}=-\partial\mathbf{A}/\partial t = a_0\left( -\mathbf{e_y}\sin\xi + \mathbf{e_z}\cos\xi \right)$ is the electric field and $\mathbf{B}=\curl{A} = a_0\left( -\mathbf{e_y}\cos\xi + \mathbf{e_z}\sin\xi \right)$ is the mangetic field.
It is seen from Eqs.~\eqref{eq.A.px} and~\eqref{eq.A.chi} that in the classical aprroach $\chi$ is constant during the particle motion. If we account QED effects then $\chi$ parameter changes after each act of gamma-quanta radiation. Although this process can be considered instantaneous, for the sake of estimating the energy loss it is convenient to introduce a continuous force of the radiation friction $F_{rr}$
\begin{align}
    \label{eq.A.dedt}
    \frac{d\varepsilon}{dt} = -\mathbf{v\cdot E} - F_{rr} v^2 , \\
    \label{eq.A.dpxdt}
    \frac{dp_x}{dt} = -[\mathbf{v\times B}]_x - F_{rr} v_x , \\
    \label{eq.A.dvdt}
    \frac{d\mathbf{v}}{dt} = -\frac{1}{\varepsilon}\left( \mathbf{E-v(vE)+v\times B} + \frac{F_{rr}}{\varepsilon^2}\mathbf{v} \right) .
\end{align}
If we assume the particle to be ultrarelativistic then Eq.~\eqref{eq.A.dvdt} does not explicitly depend on the radiative losses as $F_{rr}/\varepsilon^2\ll1$. Although due to radiative losses $\varepsilon$ changes differently in time as compared to the case without radiative losses, we will assume that the equity $\mathbf{v\cdot E = [v\times B]}_x$, which follows from the conservation of the $\gamma-p_x$, still holds. Then the equation for the $\chi$ can be significantly simplified
\begin{equation}
    \label{eq.A.dchi}
    \begin{split}
        \frac{d\chi}{dt}=\frac{a_0}{\mathcal{E}_S}\left( \frac{d\varepsilon}{dt} - \frac{dp_x}{dt} \right) \approx -\frac{a_0}{\mathcal{E}_S}(v^2-v_x)F_{rr} \approx \\
        \approx -\frac{a_0}{\mathcal{E}_S}(1-v_x)F_{rr},
    \end{split}
\end{equation}
where $F_{rr}\equiv I_{rad}(\chi)$ and $I_{rad}$ is defined in~\eqref{eq.Irad0} and expression for which can be found, for example, in Ref.~\cite{BaierKatkov}. We assume again that the term $v_x$ can be calculated according to the classical approach without accounting radiative losses, i.e.
\begin{equation}
    \begin{split}
        1-v_x = 1 - \frac{p_x}{\varepsilon} = \frac{\varepsilon - p_x}{\varepsilon} = \frac{\rho_0}{\varepsilon} = \\
        = \frac{\rho_0}{\varepsilon_0+{\left( \frac{9}{2} a_0^2 \rho_0 t^2  \right)}^{1/3}} .
    \end{split}
\end{equation}
So finally
\begin{equation}
    \frac{d\chi}{dt}=-\frac{\chi_0 I_{rad}(\chi)}{\varepsilon_0+{\left( \frac{9}{2} a_0^2 \rho_0 t^2  \right)}^{1/3}} .
\end{equation}
The total energy loss due to radiation then is calculated as follows
\begin{equation}
    \label{eq.A.DGrad}
    \Delta\varepsilon_{rad} = \int\limits_0^{\Delta t}I_{rad}(\chi(t))dt\approx\int\limits_0^{\Delta x}I_{rad}(\chi(t))dt.
\end{equation}

\begin{figure}
    \includegraphics[width=85mm]{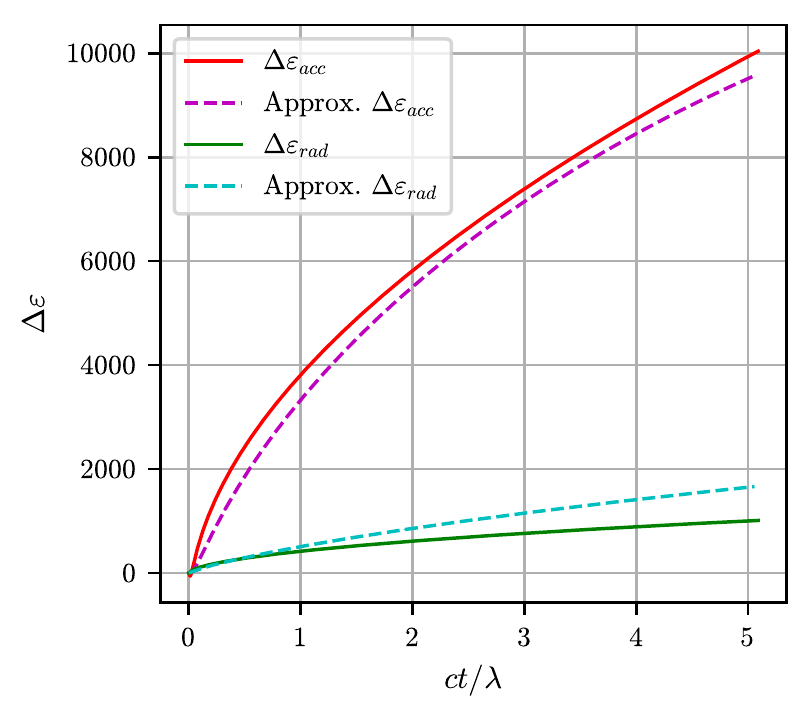}
    \caption{\label{fig.app.G} Validation of the approximations used to describe energy gain due to acceleration and energy loss due to gamma-quanta emission by a single particle in the plane wave plotted for the $a_0=2500$, $p_0=500$, $\theta=\pi$. The solid lines correspond to the numeric solution of Eqs.~\eqref{eq.A.dedt}--~\eqref{eq.A.dvdt}; the dashed lines correspond to the approximations~\eqref{eq.A.eacc} and~\eqref{eq.A.erad},where $\mu=ct/3\lambda$.}
\end{figure}

The values of $\Delta\varepsilon_{acc}$ and $\Delta\varepsilon_{rad}$ obtained using this approach overestimate the corresponding values calculated numerically from the solution of the particle motion equations with the account for the radiation friction~[Eqs.~\eqref{eq.A.dedt} and~\eqref{eq.A.dvdt}], though the order of magnitude is estimated correctly. In the cascade model the value $\Delta x$ is unknown because it is the distance along the $x$-axis between the position of the particle birth and the position where the particle crosses the moving cascade front. This quantity  cannot be calculated in our model. Also we are interested only in the total energy change during the time the particle stays in the vacuum region, so the exact dependency of the particle energy on the time is irrelevant. Thus we calculate the values $\Delta\varepsilon_{acc}$ and $\Delta\varepsilon_{rad}$ as follows
\begin{align}
    \label{eq.A.eacc}
    \Delta\varepsilon_{acc} = \mu 2^{1/3} a_0^{2/3} \varepsilon_0^{1/3} \left( 1-\cos\theta \right)^{1/3}, \\
    \label{eq.A.erad}
    \Delta\varepsilon_{rad} = \int\limits_0^{\sqrt{4\mu^3/9}} I_{rad}(\chi)dt, \\
    \frac{d\chi}{dt}=-\frac{\chi_0 I_{rad}(\chi)}{\varepsilon_0+{\left[ \frac{9}{2} a_0^2 \varepsilon_0 t^2 (1-\cos\theta) \right]}^{1/3}} .
\end{align}
where $\mu$ is the fitting parameter which determines the characteristic time the particle spends in the vacuum region. We set this parameter equally for all the particles. It turns out that by tuning this parameter both the values of $\Delta\varepsilon_{acc}$ and $\Delta\varepsilon_{rad}$ can be estimated with good accuracy (see Fig~\ref{fig.app.G}). 

\section{Effective dielectric permittivity of the $e^+e^-$ plasma}
\label{app.Electrodynamics}
\begin{figure}
	\includegraphics[width=85mm]{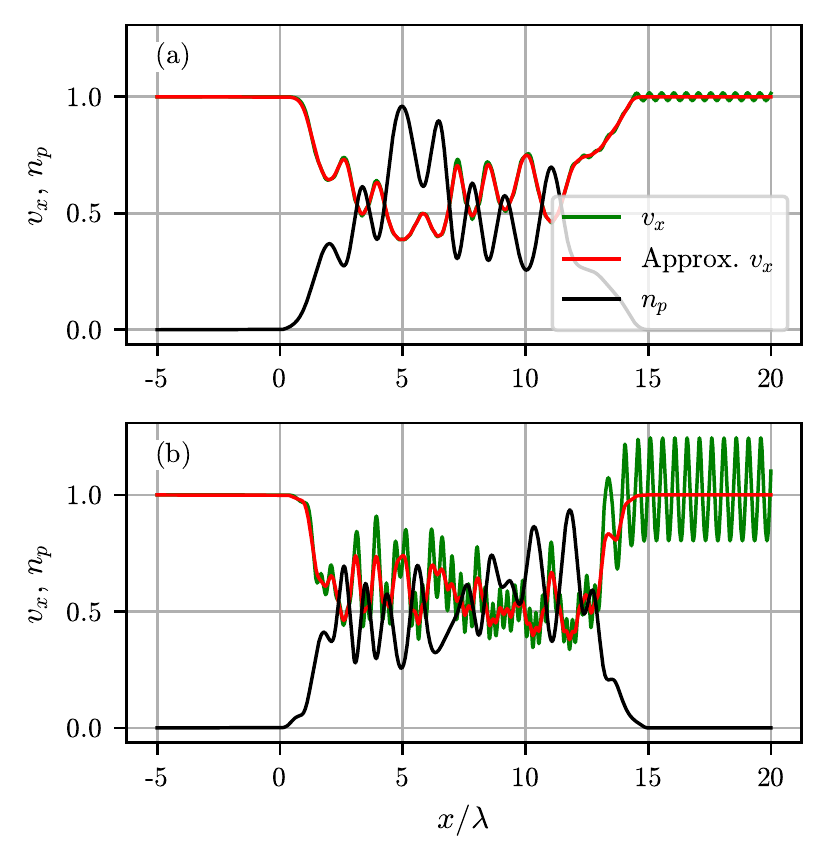}
	\caption{\label{fig.stationary} The longitudinal velocity $v_x$ computed from numerical solution of Eqs.\eqref{eq.B.Cushion1}--\eqref{eq.B.Cushion2} (green line) and from the approximate expression~\eqref{eq.B.vx} (red line) for the plasma density in the form of a inhomogeneous slab (black line). The inhomogeneity scale is smaller than the wavelength hence the WKB approximation is valid in (a) and The inhomogeneity scale is larger than the wavelength in (b).}
\end{figure}

Let us consider the propagation of a circularly polarized plane electromagnetic wave along the $x$-axis inside a medium, inhomogeneous along the $x$-axis, which induces the current $\mathbf{j}=2n_p\mathbf{v}$. The Maxwell's equations take the form
\begin{align}
    \label{eq.B.Maxwell1}
    &\frac{\partial E_z}{\partial x} = \frac{\partial B_y}{\partial t}, \\
    \label{eq.B.Maxwell2}
    &\frac{\partial E_y}{\partial x} = -\frac{\partial B_z}{\partial t}, \\
    \label{eq.B.Maxwell3}
    &\frac{\partial B_z}{\partial x} = -\frac{\partial E_y}{\partial t} - 2 n_p v_y, \\
    \label{eq.B.Maxwell4}
    &\frac{\partial B_y}{\partial x} = \frac{\partial E_z}{\partial t} + 2 n_p v_z. \\
\end{align}
It is convenient to introduce the following complex variables
\begin{align}
    &\epsilon=E_y+iE_z, 
    \label{eps} \\
    &\beta=B_z-iB_y, \\
    &v_y+iv_z=\frac{\epsilon}{|\epsilon|}\left( v_E+iv_{E \perp} \right) ,
\end{align}
where $v_E$ and $v_{E\perp}$ are the plasma velocities along and across the electric field respectively. We may also introduce the vector potential $a$ as follows
\begin{equation}
    \epsilon=-\frac{\partial a}{\partial t} \text{, } \beta=\frac{\partial a}{\partial x}.
\end{equation}
As a result Eqs.\eqref{eq.B.Maxwell1}--\eqref{eq.B.Maxwell4} can be rewritten in the form
\begin{equation}
    \frac{\partial^2 a}{\partial x^2}=\frac{\partial^2 a}{\partial t^2}-2 n_p \frac{\partial a}{\partial t}\left| \frac{\partial a}{\partial t} \right|^{-1} \left( \overline{v}_E+i\overline{v}_{E\perp} \right).
\end{equation}
We seek the solution in the form of a plane monochromatic wave with a variable amplitude
\begin{equation}
    a=E(x)e^{i\int^{x}\kappa(x)dx-it},
    \label{a}
\end{equation}
where both $E(x)$ and $\kappa(x)$ are real functions and $E(x)$ is the amplitude of the electric field. The final form of the equations is
\begin{align}
    \label{eq.B.Cushion1}
    \frac{\partial^2 E}{\partial x^2}+E(1-\kappa^2)+2 n_p v_{E\perp} =0, \\
    \label{eq.B.Cushion2}
    E \frac{\partial \kappa}{\partial x} + 2\kappa \frac{\partial E}{\partial x} - 2 n_p v_E = 0.
\end{align}
If the plasma is slightly inhomogeneous then we can apply the WKB approximation to solve the problem. If the plasma density distribution has a form of a inhomogeneous slab (as in the case of the cascade development) then this approximation is valid inside the plasma and may be invalid near the edges. Assuming the scale of the plasma inhomogeneity $L$ is larger than the laser wavelength $\lambda$ we can neglect the term with the second derivative: $\partial^2E/\partial x^2 \sim E/L^2 \ll k^2 E = (2\pi)^2E/\lambda^2$. So
\begin{equation}
    E(1-\kappa^2)+2 n_p v_{E\perp} = 0.
\end{equation}
Solving this equation we get
\begin{equation}
    \label{eq.B.kappa}
    \kappa \equiv \frac{B}{E}=\sqrt{1+\frac{2 n_p v_{\perp}}{E}},
\end{equation}
where we $v_\perp^2=v_E^2+v_{E\perp}^2=v_y^2+v_z^2$. We specify the expression $v_\perp$ as follows [see Eq.~\eqref{eq.j}]
\begin{equation}
v_\perp = \nu \sqrt{1-v_x^2}.
\label{vp}
\end{equation}
Note that in the case $\nu>0$ according to~\eqref{eq.B.kappa} $B>E$ and thus $1/\kappa$ has a meaning of the drift velocity $v_x$, so
\begin{equation}
    \frac{1}{v_x}=\sqrt{1+\frac{2n_p\nu}{E}\sqrt{1-v_x^2}} .
\end{equation}
The solution of that equation is the following~\cite{Samsonov2019}
\begin{align}
    \label{eq.B.vx}
    &v_x= {\left( \frac{2}{1+\sqrt{1+S^2}} \right)}^{1/2},\\
    &S=\frac{4n_p\nu}{E}.
\end{align}
The comparison of this approximate expression with the exact numerical solution of Eqs.\eqref{eq.B.Cushion1}--\eqref{eq.B.Cushion2} is shown in Fig.~\ref{fig.stationary} for both cases whether the WKB approximation is valid or not.

Since $v^2=1$ is assumed in the derivation of Eq.~\eqref{eq.B.vx} it is valid for any reference frame where particles are ultrarelativistic, e.g. the laboratory reference frame.

\bibliography{main.bib}

\end{document}